\documentclass[conference]{IEEEtran}
\IEEEoverridecommandlockouts
\usepackage{cite}
\usepackage{amsmath,amssymb,amsfonts}
\usepackage{algorithmic}
\usepackage{graphicx}
\usepackage{textcomp}
\usepackage{xcolor}
\usepackage{listings}
\usepackage{color}
\usepackage{soul}
\usepackage{subfigure}
\def\BibTeX{{\rm B\kern-.05em{\sc i\kern-.025em b}\kern-.08em
    T\kern-.1667em\lower.7ex\hbox{E}\kern-.125emX}}
\begin{document}

\title{A Byzantine Fault-Tolerant Consensus Library for Hyperledger Fabric}

\author{\IEEEauthorblockN{Artem Barger}
\IEEEauthorblockA{\textit{IBM Research}\\
Haifa, Israel \\
bartem@il.ibm.com}
\and
\IEEEauthorblockN{Yacov Manevich}
\IEEEauthorblockA{\textit{IBM Research}\\
Haifa, Israel \\
yacovm@il.ibm.com}
\and
\IEEEauthorblockN{Hagar Meir}
\IEEEauthorblockA{\textit{IBM Research}\\
Haifa, Israel \\
hagar.meir@ibm.com}
\and
\IEEEauthorblockN{Yoav Tock}
\IEEEauthorblockA{\textit{IBM Research}\\
Haifa, Israel \\
tock@il.ibm.com}
}

\newcommand{\BFTSmart}{\textsc{BFT-SMaRt} }
\newcommand{\compF}[1]{\textbf{\textsf{#1}}}

\IEEEoverridecommandlockouts
\IEEEpubid{\makebox[\columnwidth]{978-0-7381-1420-0/21/\$31.00~\copyright2021 IEEE \hfill} \hspace{\columnsep}\makebox[\columnwidth]{ }}

\maketitle

\IEEEpubidadjcol

\begin{abstract}

Hyperledger Fabric is an enterprise grade permissioned distributed ledger platform that offers modularity for a broad set of industry use cases. One modular component is a pluggable ordering service that establishes consensus on the order of transactions and batches them into blocks. However, as of the time of this writing, there is no production grade Byzantine Fault-Tolerant (BFT) ordering service for Fabric, with the latest version (v2.3) supporting only Crash Fault-Tolerance (CFT).

In this work we describe the design and implementation of a BFT ordering service for Fabric, employing a new BFT consensus library. The new library, based on the \BFTSmart protocol and written in Go, is tailored to the blockchain use-case, yet is general enough to cater to a wide variety of other uses. The BFT library's design and integration into Fabric address crucial aspects that were left unsolved in all prior work, making them unfit for production use. We evaluate the new BFT ordering service by comparing it with the currently supported Raft-based CFT ordering service in Hyperledger Fabric.

\end{abstract}

\begin{IEEEkeywords}
Blockchain, Distributed Ledger
\end{IEEEkeywords}

\section{Introduction}

Hyperledger Fabric (or just Fabric) is an open source project dedicated to the development
of an enterprise grade permissioned blockchain platform ~\cite{fabric-home}.
Fabric employs the \emph{execute-order-validate} paradigm for distributed execution of smart contracts.
In Fabric, transactions are first tentatively \emph{executed}, or endorsed, by a subset of peers.
Transactions with tentative results are then grouped into blocks and \emph{ordered}.
Finally, a \emph{validation} phase makes sure that transactions were properly endorsed and are not in
conflict with other transactions. All transactions are stored in the ledger, but valid transactions
are then committed to the state database, whereas invalid transactions are omitted from the state \cite{FabricPaper}.

In the heart of Fabric is the ordering service, which receives endorsed
transactions from the clients, and emits a stream of blocks.
At the time of writing, the latest Fabric release (v2.3) uses the Raft ~\cite{Raft-ATC14,Raft-github} protocol which is Crash Fault Tolerant (CFT). Despite previous efforts to do so, Fabric still does not have a Byzantine Fault-Tolerant (BFT) ordering service.
In this paper we describe our efforts to transform Fabric into a end-to-end BFT system, and contrast our approach with previous attempts.

The latest attempt to provide Fabric with a BFT ordering service, by Sousa et al. in 2018 \cite{BFTSmartFabric},
adapted the Kafka-based ordering service of Fabric v1.1 and replaced Kafka with a cluster of
\BFTSmart servers. That attempt was not adopted by the community \cite{mailing-list-fabric-on-bft}
because of various reasons (elaborated in Section~\ref{sec:related}). The main problem was that it was built on the Kafka architecture and used an external ``general-purpose'' monolith BFT cluster. The implications are that it did not address some of the more difficult, configuration related work flows of Fabric, and that the BFT cluster followers were not able to validate the transactions proposed by the leader during the consensus phase. In addition, it missed out opportunities to perform several much needed optimizations that exploit the unique blockchain use-case. Moreover, a two process two language solution increases the complexity and cost of deploying, operating and maintaining the service.

In the time that passed since then, Fabric incorporated the Raft protocol as the core of the ordering service, significantly changing the Ordering Service Node (OSN) in the process. Our goal was to implement a BFT library in the Go programming language, that would be fit to use as an upgrade for Raft.
Our prime candidates were \BFTSmart \cite{BFTSmart} and PBFT \cite{PBFT}. We soon realized that simply re-writing the Java (or C) implementation in Go will not make the cut. Fabric itself presents some unique requirements that are not addressed by traditional protocol implementations. In addition, the blockchain use case offers many opportunities for optimizations that are absent from a general-purpose transaction-ordering reference scenario. We therefore set out to design and implement a BFT library that on the one hand addresses the special needs and opportunities of a blockchain platform, but on the other hand is general and customizable enough to be useful for other use cases.

One of our goals was to provide an end-to-end Fabric system that addresses all the concerns that a BFT system must face. This forced us to tackle issues that span the entire Fabric transaction flow: the client, the ordering service, and the peers. The result is the first fully functional BFT-enabled Fabric platform.
Our key contributions are:
\begin{itemize}
  \item A stand-alone Byzantine fault-tolerant consensus \textbf{library}, based on \BFTSmart. The code is open source \cite{LibraryOpenSource} and is written in the Go programming language.
  \item An easy to integrate \textbf{interface} of a consensus library, suited for blockchains. This interface
      captures the special needs of a blockchain application, but is fully customizable for other use cases as well.
  \item A full \textbf{integration} of the library with Hyperledger Fabric~\cite{FabricPaper}, which addresses BFT concerns of all its components. This BFT version of Fabric is publicly available and open source \cite{FabricBFTOpenSource}, with an accompanying SDK \cite{SDKBFTOpenSource-Java}.
  \item An \textbf{evaluation} of our BFT version of Fabric versus Fabric based on Raft. The evaluation demonstrates that our implementation is comparable in performance to the  earlier \BFTSmart based implementation \cite{BFTSmartFabric}, but slower than Raft, mainly due to the lack of pipelining.
\end{itemize}

The library and BFT-Enabled Fabric system presented in this work form the basis of an industrial asset tokenization platform (see press releases \cite{PR-1,PR-2}), and is currently a candidate Fabric RFC \cite{FabricRFC}, discussed for inclusion in a future Fabric release \cite{FabricRFC-BFT}.

The rest of the paper is organized as follows. Section~\ref{sec:background} introduces some background on consensus protocols and Hyperledger Fabric. In Section~\ref{sec:architecture} we provide a high level view of Fabric's new BFT Ordering Service Node, describing the main internal components and their relation to the library. Section~\ref{sec:library} provides a more detailed exposition of the BFT consensus library we developed, whereas Section~\ref{sec:fabric} describes the additions and modifications we had to perform to Fabric's orderer, peer, and client, in order to turn it into an end-to-end BFT system. In Section~\ref{sec:evaluation} we evaluate the fully integrated BFT Fabric system, and compare it to the current Raft-based implementation. Finally, Sections \ref{sec:related} and \ref{sec:conclusion} discuss related work and summarize our conclusions, respectively. 

\section{Background} \label{sec:background}

\subsection{Consensus}
Most enterprise blockchain platforms use quorum-based consensus protocols to order transactions. Loosely speaking, crash fault tolerant (CFT) protocols like Raft~\cite{Raft-ATC14} need a simple majority $Q=\frac{1}{2}N$, whereas ``mainstream'' Byzantine fault tolerant (BFT) protocols like PBFT~\cite{PBFT} require a $Q=\frac{2}{3}N$ majority ($N$ size of cluster, $Q$ size of quorum). PBFT is more expensive than Raft: it requires more replicas to protect against the same number of faults ($F=N-Q$), executes an additional communication round (3 vs. 2), and requires the use of cryptographic primitives. The benefit is protection against arbitrary faults, including malicious behavior and collusion. If the number of faults is no more than $F$, both protocols provide \emph{safety} (all correct replicas agree on the same sequence of valid values) and \emph{liveness} (if a correct node delivers a value, then eventually all other correct nodes will deliver it as well).

\subsection{Hyperledger Fabric} \label{sec:background-fabric}
The Fabric blockchain network ~\cite{FabricPaper} is formed by nodes which could be classified into three categories based on their roles:
(1) \emph{Clients} are network nodes running the application code, which coordinate transaction execution. Client application code typically uses the Fabric SDK in order to communicate with the platform.
(2) \emph{Peers} are platform nodes that maintain a record of transactions using an append-only ledger and are responsible for the execution of the chaincode (smart contracts) and its life-cycle. These nodes also maintain a ``state'' in the form of a versioned key-value store. Not all peers are responsible for execution of the chaincode, but only a subset of peers called \emph{endorsing peers} \cite{EndorsmentFabric-BC19,ServiceDiscovery-IBMJRD19}.
(3) \emph{Ordering nodes} are platform nodes that form a cluster that exposes an abstraction of atomic broadcast in order to establish total order between all transactions and to batch them into blocks.

In order to address scalability and privacy, Fabric introduces the concept of channels. A channel in Fabric allows a well defined group of organizations that form a consortium to privately transact with each other. Each channel is essentially an independent private blockchain, with its own ledger, smart contacts, and a well defined set of participants.

Fabric's transaction flow is:
(1) A client uses an SDK to form and sign a transaction proposal, and sends the transaction proposal to a set of endorsing peers.
(2) Endorsing peers simulate the transaction by invoking the chaincode, and send signed endorsements back to the client.
(3) The client collects responses from all endorsing peers, validates their conformance, and packs the responses creating a transaction.
(4) The client then submits the transaction to the ordering service.
(5) The ordering service collects incoming transactions, packs them into blocks, and then orders the blocks to impose total order of transactions within a channel context.
(6) Blocks are delivered to all the peers, for example by peers pulling blocks directly from the ordering service.
(7) Upon receiving a new block, a peer iterates over the transactions in it and validates: a) the endorsement policy, and b) performs multi-version concurrency control checks.
(8) Once the transaction validation has finished, the peer appends the block to the ledger and updates its state.

Fabric is a modular blockchain system supporting multiple types of ordering services.
In Fabric's first release (v1.0) the ordering service was based on Kafka \cite{Kafka}, a replicated, (crash) fault tolerant messaging platform. The ordering service nodes (OSNs) sent transactions to a Kafka topic (one for each channel), and consumed from it an ordered transaction stream. Then the OSNs employed a deterministic function to partition
the transactions into identical blocks across all nodes. In this architecture the ordering nodes did not communicate between them directly; they only acted as producers and consumers of a Kafka service. Moreover, every node was servicing all the channels.

On release v1.4.1 Fabric introduced an ordering service based on a Go implementation of the Raft \cite{Raft-ATC14} consensus algorithm from \emph{etcd} \cite{ETCD}. This significantly changed the architecture of the OSN. Each channel now operates an independent cluster of Raft nodes. An OSN can serve multiple channels, but is not required to serve all of them. This permits linear scalability in the number of channels by spreading channels across OSNs. 
Raft is a leader-based protocol. The leader of each cluster (channel) batches incoming transactions into a block, and then proposes that block to the consensus protocol. The result is a totally ordered stream of blocks replicated across all OSNs that serve the channel. Clients are required to submit transactions to a single node, preferably the leader. However, transactions may be submitted to non-leader nodes, which then forward them to the leader. At the time of this writing Fabric's latest version (v2.3) offers the Raft-base ordering service as default, and deprecated the Kafka-base option.

\subsection{The \textsc{BFT-SMaRt} Protocol}
Parts of the library presented in this paper were designed based on the \textsc{BFT-SMaRt} consensus library~\cite{BFTSmart,BFTSmartTransformation}. In \textsc{BFT-SMaRt} the message pattern in the normal case is similar to the PBFT protocol~\cite{PBFT}, i.e. \textsc{pre-prepare}, \textsc{prepare}, and \textsc{commit} phases/messages.
If the current leader is faulty, a new leader is elected using a \emph{view change} protocol, which we implement with the \emph{synchronization} phase of \textsc{BFT-SMaRt}~\cite{BFTSmartTransformation} in mind.
The view change process uses three types of messages: \textsc{view-change}, \textsc{view-data}, and \textsc{new-view}.
We chose to implement the \textsc{BFT-SMaRt} protocol because of its simplicity and elegance. This protocol is significantly simpler than PBFT, because it does not allow for a transaction pipeline. In \textsc{BFT-SMaRt} there is only a single proposed transaction by a given leader at any point in time, which dramatically simplifies the view change sub-protocol.

\section{Architecture} \label{sec:architecture}
\begin{figure*}[htbp]
\centering
\includegraphics[width=0.9\linewidth]{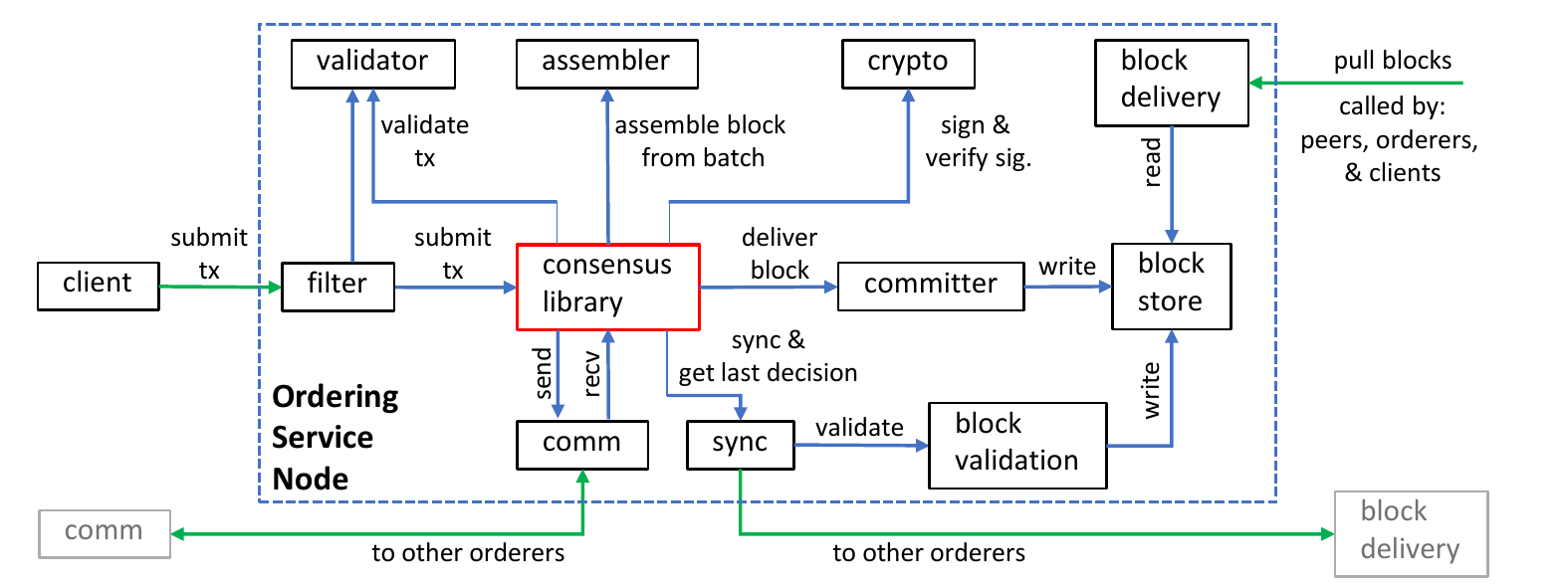}
\caption{The architecture of a Fabric BFT-based Ordering Service Node.}
\label{fig:architecture}
\end{figure*}

We now describe the architecture of the new BFT OSN, following the main components depicted in Figure~\ref{fig:architecture}. We present the novel library interface in Sec.~\ref{sec:library}, and the BFT related improvements to other Fabric components in Sec.~\ref{sec:fabric}.

After a client collects endorsements and assembles a transaction, it submits it to all the OSNs (see Sec.~\ref{sec:fabric-client}). An OSN will first \compF{filter} incoming transactions using several rules encoded in the \compF{validator}; for example, verifying the client's signature against the consortium's definitions. Valid transactions are then submitted to the BFT \compF{consensus library} for ordering. Submitted transactions are queued in a request pool within the library.

The consensus library is leader based, and at any given moment the node will operate either as a leader or a follower.
A leader OSN will batch incoming transactions, and then call the \compF{assembler} to assemble these transactions into a block.
The \compF{assembler} is in charge of ensuring that block composition adheres to the rules and requirements of Fabric, and returns a block that may contain some or all of the transactions in the batch, and may even reorder them. Transactions that were not included in the block will be included in a later batch. The leader will then propose the block, starting the consensus protocol. A follower node will receive this block proposal from the \compF{comm} component and will revalidate every transaction in it using the same \compF{validator} used by the \compF{filter}. This is done because in a BFT environment the follower cannot trust the leader -- it must therefore validate the proposal against the application running on top of it -- the Fabric blockchain. After a block proposal passes through the first two phases of consensus the \emph{commit} message is signed using the \compF{crypto} component and sent to all the nodes. Every node that receives a commit uses the \compF{crypto} component to verify the signature on it. When enough valid commit messages accumulate, the block along with all the $Q=2F+1$ commit signatures is delivered to the \compF{committer} component that saves it into the \compF{block store}. We call the block and commit signatures the ``decision''.

Each OSN has a \compF{block delivery} service that allows peers, orderers and clients to request blocks from it. This is how peers get their blocks, and this is how an OSN that was left behind the quorum catches up. If the consensus library suspects that it is behind the frontier of delivered decisions, it will call the \compF{sync} component, which in turn uses the \compF{block delivery} service of other OSNs in order to pull the missing blocks. Every block that it receives goes through a \compF{block validation} component that makes sure it is properly signed by $Q$ OSNs, as required by the BFT protocol. When it finishes, it will provide the library with the most recent decision.

This architecture resembles the Raft-based OSN architecture in the sense that the consensus library is embedded within the OSN process. However, there are important differences:
\begin{itemize}
  \item In a Raft-OSN the block is cut before ordering and is then submitted to the raft library for ordering. Here transactions are submitted to the BFT library, because the library monitors against transaction censorship. This required the addition of the \compF{assembler} to the OSN and respective APIs to the library.
  \item In a Raft-OSN the followers trust the leader's block proposal (as it can only crash fail), whereas here the followers must revalidate the transactions within the block proposal. This required adding to the library APIs for transaction verification during the consensus protocol.
  \item In a Raft-OSN the delivered block is signed after consensus by the node, right before it is saved to the block store. Here the block is signed during the consensus protocol, in the commit phase, by $Q$ nodes.
\end{itemize}

\section{The consensus library and API} \label{sec:library}

We implemented a BFT consensus algorithm in a stand-alone open-source library \cite{LibraryOpenSource}.
The algorithm is mostly based on the well known PBFT \cite{PBFT, PBFTrecovery} and \BFTSmart \cite{BFTSmart, BFTSmartTransformation} protocols. However, the library interface differs in several important ways from the traditional BFT library API.

\subsection{Request life-cycle}
\begin{figure*}[htb]
	\centering
	\includegraphics[width=\linewidth]{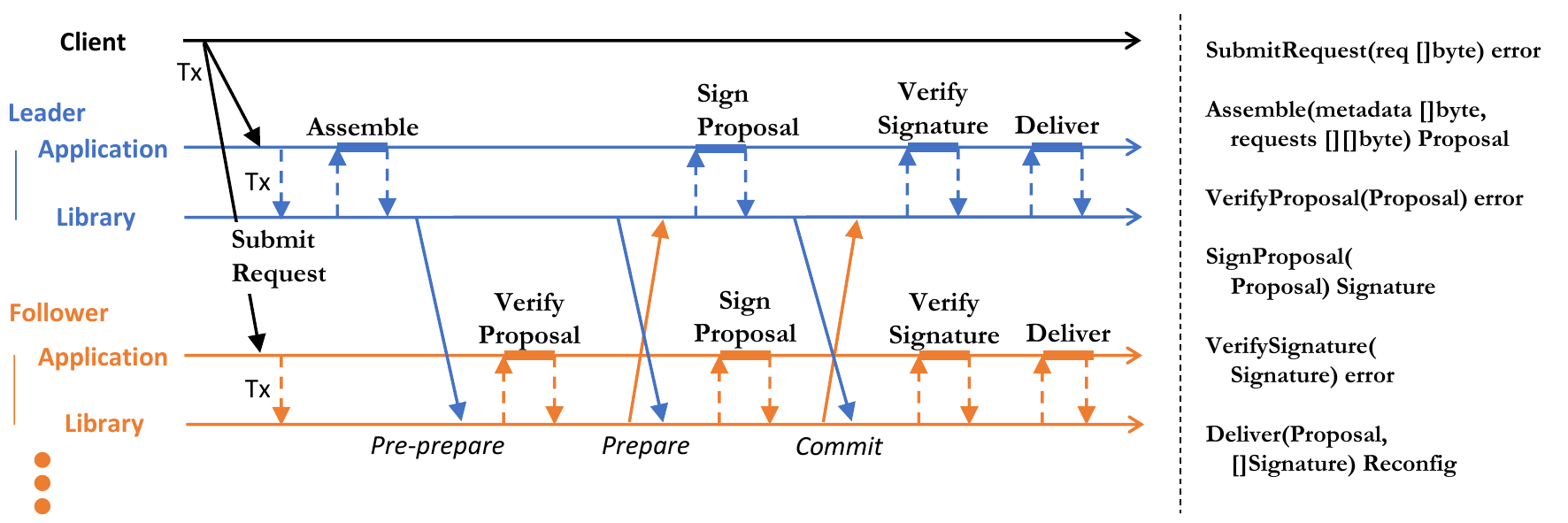}
	\caption{Normal case library-application flow (left), and API (right).}
	\label{fig:lib-flow}
\end{figure*}

As in most consensus protocols, the request life-cycle, shown in Fig.~\ref{fig:lib-flow}, starts with ``submit'' and ends with ``deliver'':

\newcommand{\funcF}[1]{\textbf{\textsf{#1}}}

\funcF{SubmitRequest(req []byte) error} -
starts the request life-cycle. Clients should submit requests to all nodes and not just to the leader. Whenever a request is not included in a proposal after a first timeout, every non-faulty follower \emph{forwards} the request to the leader. This protects against malicious clients sending messages only to followers, trying to force a view change. If a request is not included in a proposal after a second timeout, every non-faulty follower will initiate a view change. This protects against faulty leaders.

\funcF{Deliver(Proposal,[]Signature) Reconfig} -
is the end of the request life-cycle. It is invoked by the library when a node receives a quorum of \textsc{commit} messages, and delivers a committed proposal (a block) along with $Q$ signatures to the application. The application is responsible to store the delivered decision, as the library may dispose of this data once the deliver call returns. The application returns a \textsc{Reconfig} object that tells the library whether any of the transactions in the \textsc{Proposal} reconfigured the library (see below).

The requirements of the Fabric OSN guided us into a library API that extends this pattern, and requires the application to implement several additional functions. The invocation flow is depicted in Figure~\ref{fig:lib-flow}. This extension makes the API perfectly suited to the blockchain domain.

\funcF{Assemble(metadata []byte, reqs [][]byte) Proposal} -
is called by the leader with a batch of requests along with some metadata (which are \textsc{sequence} and \textsc{view-number}). This allows a blockchain application to construct a block (\textsc{Proposal}) in any form it wishes. For example, this enables adding block and chain specific information, such as hash of the previous block, merkle-tree, etc. Moreover, in Fabric, a configuration transaction must reside in a block by itself.

\funcF{VerifyProposal(Proposal) error} -
is called by a follower that receives a proposal in a \textsc{pre-prepare} message from the leader (since the leader may be malicious). In Fabric, we check that the proposal is constructed as a valid and correctly numbered block, its header contains the hash of the previous block, and all transactions in the block are valid Fabric transactions.

\funcF{SignProposal(Proposal) Signature} -
is called during the \emph{commit} phase when nodes sign the current \textsc{Proposal} and send the signature as part of the \textsc{commit} message. Nodes also sign the \textsc{view-data} message during the view change process (not shown).

\funcF{VerifySignature(Signature) error} -
is called to verify incoming \textsc{commit} messages, as part of the \emph{commit} phase, where each node collects a quorum of \textsc{commit} messages.

\funcF{Sync() Reconfig} -
is called by the library when it suspects it is behind by several commits (e.g. recovering after a long down-time). This will trigger a BFT-compliant replication protocol (see Sec. ~\ref{sec:fabric}) and pull the missing blocks. The library will be provided with a \textsc{Reconfig} object that tells it whether any of the transactions in the pulled history reconfigured the library (see below).

\funcF{VerifyRequest(req []byte) error} -
The library assumes the application submits only valid client requests. However, there are cases where a client's request must be reverified:
(1) if a follower forwards a request to the leader, the leader calls \funcF{VerifyRequest};
(2) if the configuration changed, as elaborated next, then all queued client requests must be reverified.

\subsection{Reconfiguration}
The library supports dynamic (and non-dynamic) reconfiguration.
Reconfiguration is done implicitly and not explicitly. Instead of the application sending an explicit reconfiguration transaction to the library, the library infers its configuration status after each commit. To that end, the  \funcF{Deliver} and \funcF{Sync} calls return a \textsc{Reconfig} object that defines a new configuration, if there was any. If there was a reconfiguration, the library restarts all of its components with the new configuration.

\subsection{Communication}
The library assumes the existence of a communication infrastructure that implements point to point authenticated channels (like Raft). The library is notified when a new message arrives; messages carry the identifier of the node that sent them.
The communication infrastructure is used to send both \emph{Consensus protocol messages} (i.e. \textsc{pre-prepare}, \textsc{prepare}, \textsc{commit}, view-change protocol messages, and heartbeats), as well as \emph{Forwarded clients' requests}.

\subsection{Write ahead log}
The WAL (write-ahead log) is a pluggable component, and we provide a simple implementation that maintains only the latest state. The WAL can be kept that simple because a blockchain application will keep a history of all delivered decisions. In contrast, a general purpose SMR may apply decisions to the state and discard them, and therefore needs to keep all decisions to the latest state snapshot.

\subsection{Summary}
The transaction flow in the monolith \BFTSmart cluster used in \cite{BFTSmartFabric} starts with \textsc{submit} and ends with \textsc{deliver}, with no means for efficiently implementing the requirements of a blockchain system. The design of our library addresses this gap by defining API functions that need to be implemented by the application using the library. This allows us to address the pain points that had failed previous attempts to implement a BFT ordering service for Fabric. In less demanding applications (e.g without blocks, no need for transaction validation) these ``hooks'' into the consensus protocol can be filled by simple ``empty'' implementations.

\section{Byzantine fault-tolerant Fabric} \label{sec:fabric}

In order to turn Fabric into an end-to-end BFT system we had to make changes that go beyond replacing the Raft consensus library with our BFT library.

\subsection{Block structure}
Fabric's block is composed of a header, a payload, and metadata. One of the metadata fields is consensus specific information. We changed the type of information in it to the \textrm{ViewID} and \textrm{Sequence} number from the consensus library.

\subsection{Signatures and validation}
Every time a Fabric node (peer or orderer) receives a block outside of the consensus process, it must validate that the block is properly signed. This happens in three different scenarios: 1) when a peer receives a block from the ordering service, 2) when a peer receives a block from another peer during the gossip protocol, and 3) when an
ordering node receives a block from another ordering node during the synchronization protocol. The default policy in Fabric is to require a single signature from a node that belongs to any of the organizations that compose the ordering service consortium. For a BFT service, this is obviously not enough.

We implemented a new \emph{BFT validation policy} that requires that the block be signed by $Q$ out-of $N$ nodes. Moreover, we require that these signatures are from the consenter set (i.e. the set of $N$ nodes) defined in the last channel configuration transaction. This means that the new validation policy we defined must be dynamically updated every time the consenter set is updated with a configuration transaction, which was previously not supported.

\subsection{The orderer node}\label{sec:fabric-orderer}
Changing the orderer node to work with our library amounts to implementing the logic that initializes the library, submits transactions to the library, and handles invocations from the library.
\emph{Synchronization} happens when an orderer node falls behind the cluster. This is done by requesting blocks from other cluster members. We changed the default synchronization protocol to tolerate Byzantine faults.
\emph{The Block assembler} is in charge of inspecting a batch of incoming transactions offered to it by the leader, and returning a block of transactions that is built in accordance with the semantics of Fabric, for example, that a configuration transaction must be packed in a block by itself.
\emph{The Signer and Verifier} implementations simply wrap the already existing capabilities of Fabric to sign and verify data.

\subsection{The peer} \label{sec:fabric-peer}
In Fabric, a peer will randomly select an ordering node and request a stream of blocks from it. In our BFT implementation, the blocks that are received from the ordering service cannot be forged because they are signed by at least $2F+1$ ordering nodes. However, receiving the stream of blocks from a single orderer (or up to $F$ orderers) might expose the peer to a censorship attack. To protect against this scenario we implemented a block delivery service that is resilient to this sort of attack.

A naive expensive solution is for the peer to ask for the block stream from $F+1$ ordering nodes. Instead, we \emph{augment the delivery service} and allow a peer to request a new kind of delivery -- a stream that contains only the header and metadata of each new block, enough to verify the signatures. The peer requests a stream of full blocks from a randomly selected orderer, and a \emph{Header-Metadata} stream from all the rest. The peer then monitors the front of delivered blocks: if $F$ or more \emph{Header-Metadata} streams are ahead of the full block stream for more than a threshold period, then the peer suspects the orderer that delivers the stream of full blocks applies censorship, and replaces it with another one.

\subsection{The client} \label{sec:fabric-client}
In the Raft CFT setting a client is only required to submit a transaction proposal to a single orderer. We modified the client in the Fabric Java SDK \cite{SDKBFTOpenSource-Java} to submit every transaction to all the orderers, in order to prevent censorship attacks by malicious orderers. This is the simplest solution to always hit the leader, which minimizes forwarding between orderers.

\section{Evaluation} \label{sec:evaluation}

We evaluated the performance of our BFT ordering service (BFT-OS) for Fabric that integrates our consensus library, and compared it with the existing state-of-the-art Raft ordering service (Raft-OS, from Fabric v2.3). Here we report a summary of our findings.

\subsection{Setup and Workload}
We consider both LAN and WAN setups, different cluster sizes (4, 7, 10 for BFT, 5, 7, 11 for Raft) and various configurations of batch sizes (100, 250, 500, 1000 transactions/block). We pre-allocate 1.4 million Fabric transactions each sized $~$4KB (typical of a transaction which contains 3 certificates of endorsing peers) and send it to the leader from 700 concurrent workers. This ensures the leader always fills blocks with as many transactions it can (resulting in block sizes $\sim$400KB, $\sim$1MB, $\sim$2MB, $\sim$4MB resp.). We used a homogeneous virtual machine specification for all nodes with a 16 core Intel(R) Xeon(R) CPU E5-2683 v3 @ 2.00GHz with 32GB RAM and SSD hard drives. In our WAN setup we spread the servers across the entire globe to maximize the latency (Dallas, London, Washington, San Jose, Toronto, Sydney , Milan, Chennai, Hong Kong, and Tokyo). The leader is always located in London. Latency to the followers is: $average=133$ms, $min=20$ms (Milan), $max=250$ms (Tokyo). Each test is repeated 10 times.


\begin{figure*}
	\centering
	\subfigure[BFT in a LAN]{\includegraphics[width=0.43\linewidth]{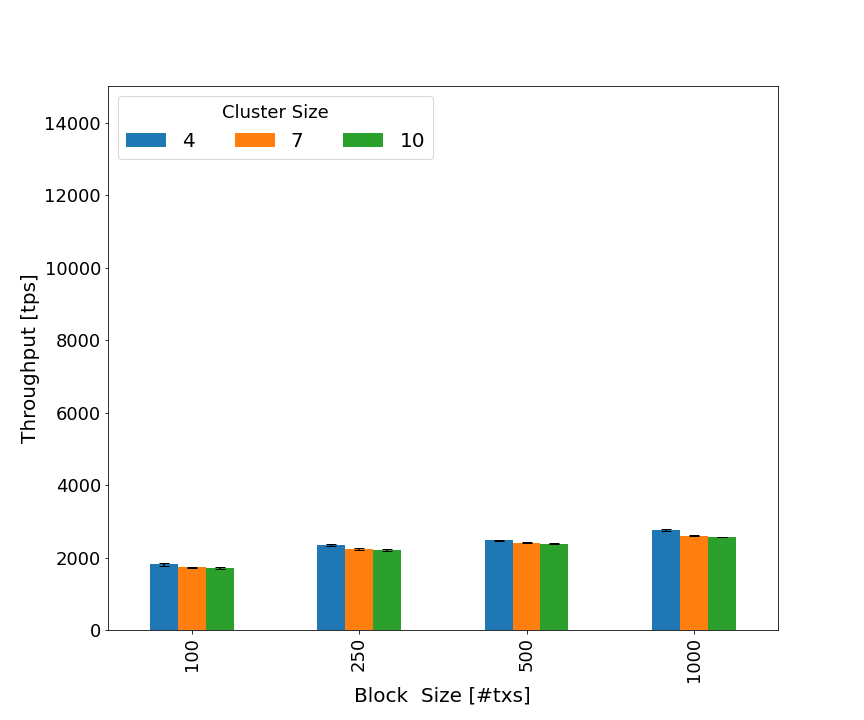}}
	\quad 
	\subfigure[Raft in a LAN]{\includegraphics[width=0.43\linewidth]{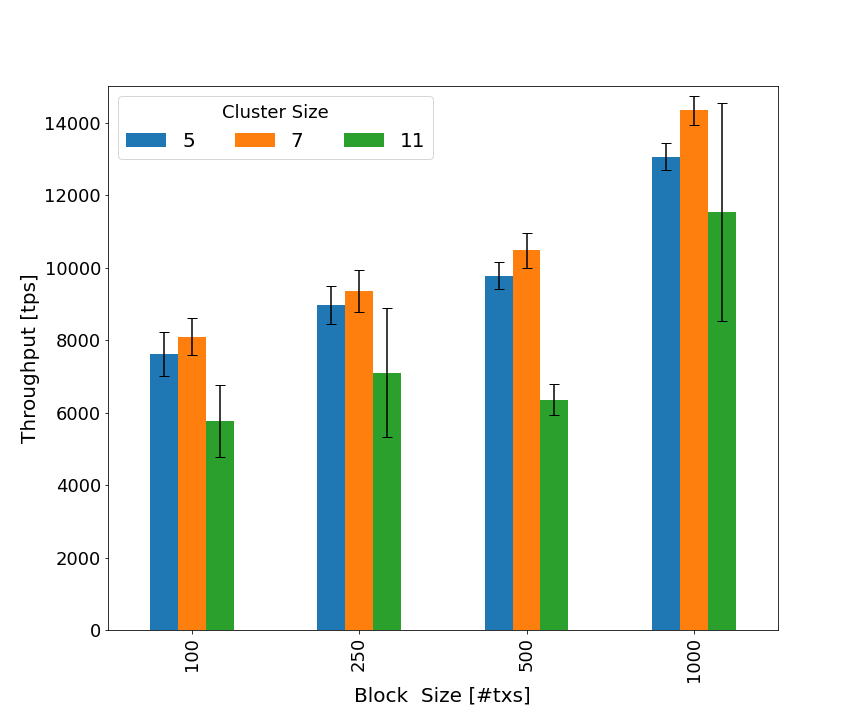}}
	\caption{Cluster throughput as a function of block size, in a LAN, for BFT (a) and Raft (b). Transaction size is $\sim$4KB.}
	\label{fig:eval-perf-lan}
\end{figure*}

\begin{figure*}
	\centering
	\subfigure[BFT in a WAN]{\includegraphics[width=0.43\linewidth]{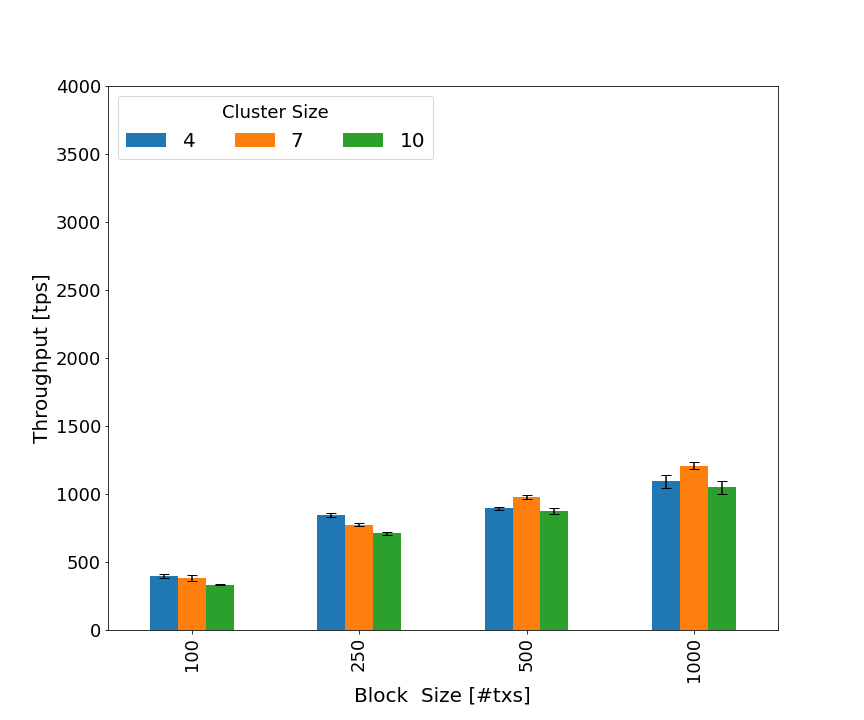}}
	\quad 
	\subfigure[Raft in a WAN]{\includegraphics[width=0.43\linewidth]{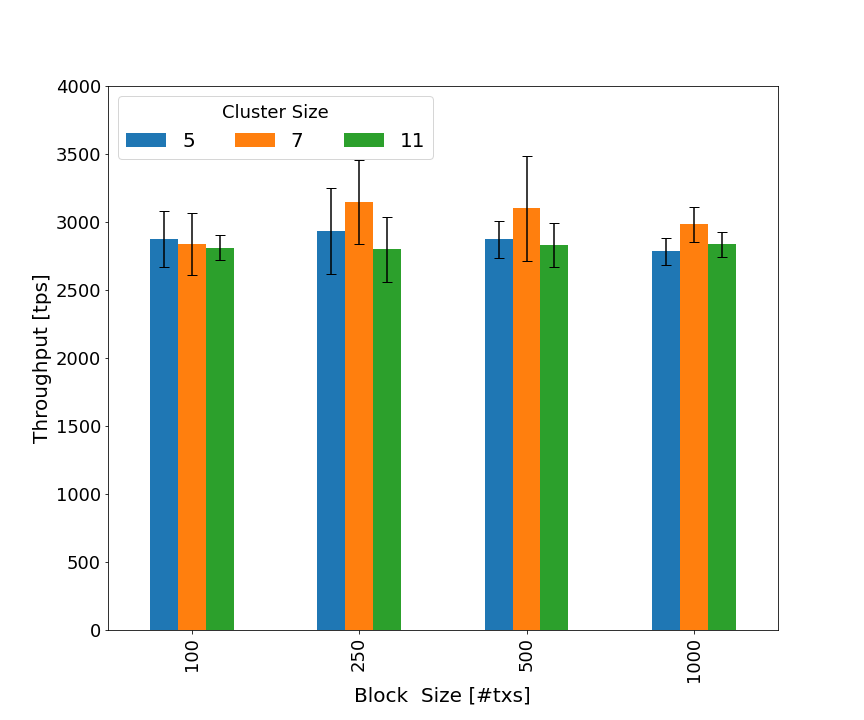}}
	\caption{Cluster throughput as a function of block size, in a WAN, for BFT (a) and Raft (b). Transaction size is $\sim$4KB.}
	\label{fig:eval-perf-wan}
\end{figure*}

%
%

\subsection{LAN performance}
Figure~\ref{fig:eval-perf-lan} depicts the throughput (in tx/sec) of both BFT-OS (left) and Raft-OS (right) with various block sizes and cluster sizes.

For both BFT-OS and Raft-OS, throughput improves with block size (i.e. the number of txs per block). For BFT-OS, a rate of $~$2,500 tx/sec means sending 80Mb of traffic per second per node, and even in a cluster of 10 nodes, it means a total fan-out of 720 Mb per second. In contrast, for Raft the situation is different, and at a throughput of 12,000 tx/sec, the leader sends 375Mb to each of the 10 followers which starts hindering the throughput.

Since the latencies in our LAN setup are negligible, we attribute the differences in throughput mainly to the high CPU processing overhead that is mandatory in BFT, but doesn't exist in Raft. We verify this conclusion by observing the CPU consumption and measuring the time it took to perform local CPU cryptographic computations for block verification, as well as monitoring I/O (omitted due to space limitations).

\subsection{WAN performance}
Figure~\ref{fig:eval-perf-wan} exhibits the results of an identical experiment, except the servers are now deployed across 10 different data centers across the globe. Throughput is significantly reduced both for BFT-OS and Raft-OS. With 1,000 tx/block, BFT-OS throughput decreases down to  40\% of the LAN setup, whereas RAFT-OS throughput decreases down to 20\% of the LAN setup.

In our BFT-OS, the leader starts consensus on a block only after the previous block has been committed. Therefore, throughput improves significantly with block size, as the WAN bandwidth is not saturated due to the lack of pipelining and high latencies. In contrast, the Raft-OS can pipeline consent on blocks, saturating the WAN bandwidth even for small blocks, and therefore does not benefit from increasing block size. The cost of increasing the block size is added average latency due to batching. In cases where the throughput is less sensitive to block size (BFT-LAN, Raft-WAN), or when latency is more important than throughput, smaller blocks may be used.

\subsection{Summary}
Using large blocks (1000 txs/block), and with resilience to two faults ($F=2$ , cluster size 7 \& 5, for BFT-OS \& Raft-OS, resp.): in a LAN, BFT-OS achieves $~$20\% the performance of a Raft-OS (2,500 vs. 13,000 tx/sec, resp.); in a WAN, BFT-OS achieves $~$40\% the performance of a Raft-OS (1,200 vs. 3,000 tx/sec, resp.). We attribute this gap mainly to the lack of pipelining in the \BFTSmart protocol, and to the cryptographic computations required in all BFT protocols.

\section{Related work} \label{sec:related}

\subsection{Blockchain and BFT}

In general, the increasing interest in blockchain lead to the development of algorithms such as Tendermint~\cite{buchman2016tendermint}, HotStuff~\cite{hotstuff} and SBFT~\cite{sbft}, where the key focus is on improving the view change sub-protocol or replacement of a faulty leader.

Many decades of research lead to BFT protocols exploiting randomization to provide consensus solution in the asynchronous model, with well known classical results such as~\cite{ben1983another,bracha1984asynchronous,toueg1984randomized,cachin2005random}. However, these protocols are far from being practical due to their poor performance. Only recently Miller et al.~\cite{HoneyBadger} suggested a leaderless randomized algorithm with reasonable and promising performance results. Recent advances include also \cite{Abraham-FC19} and \cite{locher-SPAA20}.

Despite the renaissance of research around BFT consensus algorithms, triggered primarily by increased interest in blockchain, there are only a few openly available implementations suitable for production-grade exploitation~\cite{pbftCode,TendermintCode,HoneyBadgerBFTCode}. Unfortunately, existing libraries lack a practical reusable interface which is flexible and generic enough to implement a BFT-enabled ordering service for Hyperledger Fabric. Any implementation based on a programming language other than Go would have exhibited the same drawbacks as the Kafka based ordering service architecture, where the consensus component will have been deployed as an external service. In the quest to develop a clustered ordering service based on an embedded consensus library, we had to write our own library.

\subsection{BFT-Smart and Hyperledger Fabric}

In its first release, the Fabric ordering service was based on the Kafka messaging service. In this implementation, the OSNs receive transactions from clients, and submit them to a Kafka topic, one topic for each Fabric channel. All OSNs would consume from the topics -- again, one topic per channel -- and therefore receive a totally ordered stream of transactions per channel. Each OSN would then deterministically cut the transaction stream into blocks.

In 2018 Sousa et al. \cite{BFTSmartFabric} made an attempt to convert this implementation into a BFT ordering service. They replaced the Kafka service with a cluster of BFT-Smart based servers, where each server consisted of a BFT-Smart core wrapped with a thin layer that allowed it to somewhat ``understand'' Fabric transactions. The proof of concept presented in the paper exhibited promising performance, but was eventually not adopted by the community. The community discussions reveal a mix of fundamental and technical reasons for that \cite{mailing-list-fabric-on-bft}.

The solution presented was composed of two processes (orderer front-end and BFT server), written in two languages (Go \& Java, resp.). That was not well received as it complicates the development, maintenance, and deployment of the ordering service. The experience gained with the Kafka-based service motivated the community to move towards a single process that embeds a consensus library, as eventually happened with the introduction of the Raft-based ordering service in Fabric v1.4.1.

There were, however, more fundamental reasons. The code that wrapped the BFT-Smart core did not include Fabric's membership service provider (the MSP), which defines identities, organizations, and certificate authorities (CAs), and includes the cryptographic tools for the validation of signatures. Therefore, the BFT cluster signatures were not compliant with Fabric's, and the identities of the BFT cluster servers were not part of Fabric's configuration. In Fabric, configuration is part of the blockchain and must be agreed upon. Incorporating the Java BFT cluster endpoints and identities (certificates \& CAs) into the configuration flow would have meant providing a Java implementation to an already highly sensitive component. This shortcoming also meant that the front-end orderers had to collect $2F+1$ signed messages from the BFT cluster servers, increasing the number of communication rounds to four.

The blocks produced by the front-end servers included only the signature of a single front-end orderer. This does not allow an observer of the blockchain (or a peer) to be convinced that the block was properly generated by a BFT service. Moreover, even if the $2F+1$ BFT cluster signatures were included in the block metadata, that does not help an observer, as the identities of said servers are not included in Fabric's configuration. Moreover, peers and clients did not have the policies that correspond to the BFT service they consumed from.

Another subtle problem with a monolithic BFT cluster is that it does not allow a follower to properly validate the transactions proposed by the leader against the semantics of Fabric -- again -- without pulling in a significant amount of Fabric's code.

BFT-Smart owes much of its performance to the internal batching of requests. However, those batches are not consistent with the blocks of Fabric, so had to be un-packed, multiplexed into different channels, and then re-packed into Fabric blocks.

We confronted these problems when we designed the library and its integration with Fabric. The interface of the library allowed us to seamlessly integrate with Fabric's MSP and configuration flow. Our implementation allows the leader to assemble blocks according to Fabric's rules, so transactions are batched once. It allows followers to validate the transactions against Fabric's semantics during the consensus protocol, and it collects the quorum signatures during the commit phase. This reduces the number of communication rounds to three, and allows the observer of a single block to validate the correctness of the BFT protocol. The corresponding block validation policy was added to the peer and orderer for that effect, and the SDK client was augmented to interact properly with a BFT service.

\subsection{Tendermint and Hyperledger Fabric}

Another option we considered was to re-use the Tendermint Core~\cite{TendermintCode} as an embedded consensus library. There is an Application Blockchain Interface (ABCI) which defines an API between the BFT protocol and state machine replication layer (application). However, the consensus protocol itself actually implements the blockchain, with batches of transactions chained together forming the ledger. The ABCI only provides an interface for the application to validate and execute transactions. That implies that using Tendermint as a consensus library in the Fabric OSN would have resulted in a ledger with Tendermint blocks in addition to the ledger maintained by Fabric.
In addition, Tendermint implements a custom peer-to-peer communication protocol inspired by the station-to-station protocol. This protocol significantly deviates from the communication model of Fabric. Unfortunately the current implementation is too tightly coupled into the Tendermint Core and would have required substantial refactoring.
Overall, the lack in configuration flexibility and the inability to replace the communication layer led us towards the decision to implement our own BFT library instead. 

\section{Conclusion} \label{sec:conclusion}

In this paper we described the design, implementation, and evaluation of a BFT ordering service for Fabric. In the heart of this implementation lies a new consensus library, based on the \BFTSmart protocol, and written in Go. The library presents a new API suited for permissioned blockchain applications, such as Fabric. It delegates many of the core functions that any such library must use to the application employing it, allowing for maximal flexibility and generality. For example, cryptographic functions, identity management, as well as point to point communication are
not embedded but are exposed through proper interfaces, to be implemented by the application using it. This allowed us to re-use some of the sophisticated mechanisms that Fabric already possessed. In order to make Fabric an end-to-end BFT system, we ensured that the peer and the client SDK interact properly with the BFT ordering service.

We chose to implement the \BFTSmart protocol because of its simplicity. This protocol is significantly simpler than PBFT, because it does not allow for a transaction pipeline. In BFTSmart
there is only a single proposed transaction by a given leader at any point in time, which dramatically simplifies the view change sub-protocol. This simplicity greatly increases our confidence in the correctness of the implementation. However, these advantages come with a cost -- reduced performance. This is especially salient when comparing against the highly mature and optimized etcd/raft library, which uses pipelining extensively. Nevertheless, our implementation exhibits levels of performance that are sufficient for most permissioned blockchain applications: a $7$ node BFT ordering service ($F=2$) can support $\sim2500$ TPS in a LAN, and $\sim1000$ TPS in a WAN. These numbers are for a single channel; a Fabric network can scale horizontally by adding channels.

\bibliographystyle{IEEEtran}
\bibliography{IEEEabrv,references}

\end{document}